\documentclass[aps,prc,twocolumn,showpacs,superscriptaddress,groupedaddress]{revtex4-1}  % for review and submission
\usepackage{mathrsfs,amsmath,amssymb,amsthm}
\usepackage{amssymb,fmtcount}
\usepackage{graphicx,epsfig,latexsym,overpic,amssymb,color}
\usepackage{threeparttable}
\preprint{\today}

\newcommand{\rr} {\boldsymbol{r}}

\begin{document}
\title{
Constraints on the neutron drip-line with the newly observed $^{39}$Na
}
\author{Q.Z. Chai}
\affiliation{
State Key Laboratory of Nuclear Physics and Technology, School of Physics,
Peking University, Beijing 100871, China
}
\author{J.C. Pei}\email{peij@pku.edu.cn}
\affiliation{
State Key Laboratory of Nuclear Physics and Technology, School of Physics,
Peking University, Beijing 100871, China
}
\author{Na Fei}
\affiliation{
State Key Laboratory of Nuclear Physics and Technology, School of Physics,
Peking University, Beijing 100871, China
}
\author{D.W. Guan}
\affiliation{
State Key Laboratory of Nuclear Physics and Technology, School of Physics,
Peking University, Beijing 100871, China
}

%\author{Min-Liang Liu}
%\affiliation{
%Institute of Modern Physics, Chinese Academy of Sciences, Lanzhou 730000,  China
%}
%\author{Fu-Rong Xu}
%\affiliation{
%School of Physics, Peking University, Beijing 100871,                      China
%}
%\affiliation{
%Institute of Theoretical Physics, Chinese Academy of Sciences, Beijing   100080,
%                                                                           China
%}
%\affiliation{Center of Theoretical Nuclear Physics, National Laboratory of Heavy
%                                           Ion Collisions, Lanzhou 730000, China
%}
%123456789 123456789 123456789 123456789 123456789 123456789 123456789 123456789
\begin{abstract}
The recently observed weakly-bound $^{39}$Na provides a stringent theoretical constraint on
the neutron drip-line. We studied the properties of drip-line nuclei around $^{39}$Na with the Hartree-Fock-Bogoliubov method and
various Skyrme interactions. We adopted the extended SkM$^{*}$$_{\rm ext1}$ parameterization  which can properly
describe two-neutron separation energies of oxygen and fluorine isotopes and deformations at the center of the ``island of inversion". Systematic calculations of drip lines of
O, F, Ne, Na, Mg, and Al isotopes have been performed. We infer that $^{42}$Mg is weakly bound and
 $^{45}$Al is less weakly bound. $^{44}$Mg and $^{47}$Al could be barely existed.
We also demonstrated the deformed halo properties of $^{39}$Na. Our studies could be valuable for
experimental explorations of drip-line nuclei in the forthcoming FRIB and other  rare-isotope beam facilities .

%1234567890123456789012345678901234567890123456789012345678901234567890123456789
\end{abstract}
%\pacs{  21.10.Re, 21.60.Cs, 21.60.Ev}
\maketitle
%123456789 123456789 123456789 123456789 123456789 123456789 123456789 123456789
\section{introduction}

The developments of new-generation rare-isotope beam facilities around the world
provide great opportunities for studies of exotic nuclei close to drip lines.
For example, FRIB is expected to be fully operational in 2022 and it is able to reach the neutron drip-line up to nuclei with charge number $Z$=40~\cite{frib}. It is important to study properties of the drip-line nuclei timely with highly accurate theoretical models to
guide forthcoming experiments.

Currently the drip lines of light nuclei up to O, F, Ne have been determined experimentally~\cite{Thoennessen}.
In a very recent experiment performed in RIKEN~\cite{Ahn2019PRL}, $^{31}$F and $^{34}$Ne are reconfirmed to be the drip-line nuclei.
Compared to expected experimental yields, the existence of heavier fluorine and neon isotopes have been excluded.
In this exciting experiment,  Ahn and colleagues have newly observed one event of $^{39}$Na~\cite{Ahn2019PRL}, indicating that $^{39}$Na is very weakly bound and is most likely to be the
drip-line of sodium. Their latest experiment has observed more events of $^{39}$Na~\cite{ahn2}.
The heaviest magnesium isotope been observed is $^{40}$Mg~\cite{Baumann2007}. Very recently the first excited state of
$^{40}$Mg has also been observed~\cite{crawford}. The heaviest aluminum isotope been observed is $^{43}$Al in 2007~\cite{Baumann2007}.
The experiments indicate that drip lines of magnesium and aluminum isotopes could be located beyond $^{40}$Mg and $^{43}$Al.

It is known that different theoretical models can have remarkable divergent predictions about the neutron drip lines~\cite{Erler2012nature}.
Therefore, the newly observed $^{39}$Na has put an important  stringent constraint on theoretical models.
Based on this constraint, it is time to give more reliable predictions on the drip-line locations around $^{39}$Na.

The exotic properties of weakly bound $^{39}$Na are also of great interests which has the magic neutron number of $N$=28.
While the neighboring isotones $^{40}$Mg is prolate deformed and $^{42}$Si is oblate deformed~\cite{crawford,42si}. $^{39}$Na is more
weakly bound than $^{40}$Mg and is a promising candidate for a novel deformed halo structure. Indeed, the
interplays between deformations, pairing, continuum and halos are of particular interests. In addition to novel static structures, our recent studies show that
  pygmy dipole resonances of weakly bound  $^{40}$Mg  can have novel compressional surface flows~\cite{wang}.

The suitable self-consistent theoretical framework for descriptions of deformed weakly bound nuclei is the coordinate-space Hartree-Fock-Bogoliubov (HFB)  method,
in which the continuum and halo extensions can be treated rather accurately~\cite{hfbax,zhangyn}.
The exact solution of HFB equations of deformed nuclei with outgoing boundary conditions are very difficult.
The Gamow shell model is suitable for weakly bound nuclei except for deformed cases~\cite{lijg}.
There are extensive studies on even-even nuclei, the studies of odd-$A$ nuclei such as sodium, fluorine, and  aluminum
are very needed. To this end, the quasi-particle blocking method has to be implemented in the HFB framework~\cite{xiong}.

In last few years, there are extensive efforts to develop highly accurate effective nuclear interactions for HFB or density functional theory calculations
of the whole nuclear chart~\cite{grasso,unedf0,bulgac,bcpm,kroea}. We recnetly optimized the extended Skyrme forces
with higher-order density dependencies in the $s$-wave channel~\cite{XXY2016PRC,ZZW2018}. We obtained highly accurate global descriptions of binding energies.
The widely used SkM$^{*}$ force is very successful for surface properties and fission barriers~\cite{Bartel1982}. But our systematic calculations
show that SkM$^{*}$ overestimated the extension of the neutron drip lines. However, the optimized extended SkM$^{*}$$_{\rm ext1}$ force without fitting fission barriers
obtained much improved descriptions of binding energies.
 The UNEDF0 is the best optimized Skyrme force
for global binding energies, with a rms of 1.45 MeV~\cite{unedf0}. With a
higher density dependent term, the rms of UNEDF0$_{\rm ext1}$ is further reduced to 1.29 MeV~\cite{ZZW2018}.

In this work, we aim to investigate the drip-line nuclei around $^{39}$Na with
the Skyrme Hartree-Fock-Bogoliubov method. To validate the existence of weakly bound $^{39}$Na,
various Skyrme forces have been adopted.
The deformed halo structures of $^{39}$Na have also been investigated.
For descriptions of odd-$A$ nuclei, the quasiparticle blocking method has been implemented.
The HFB calculations are performed on the axially symmetric coordinate space mesh.
The shape coexistence properties have been studied using the deformation constrained calculations.
We hope our results are useful to guide forthcoming experiments in new generation RI beam facilities
such as FRIB.

\section{The theoretical framework}
%%%%%%%%%%%%%%%%%%%%%%%%%%%%%%%%%%%%%%%%%%%%%%%%%%%%%%%%%%%%%%%%%%%%%%%%%%%%%%%%
%123456789 123456789 123456789 123456789 123456789 123456789 123456789 123456789

The Skyrme-HFB equation is solved by the HFB-AX
code~\cite{hfbax} within a two-dimensional coordinate space by
using the B-spline techniques~\cite{hfbax,Teran2003}. With large coordinate boxes and small lattice spacings,
the continuum and the halo structures can be precisely described~\cite{zhangyn}.
In this work, the box size takes 21 fm and the maximum  mesh spacing is 0.6 fm, which is
sufficient for the medium-mass region.
The HFB equation in the coordinate-space representation can be
written as~\cite{hfbax}:
\begin{equation}
    \left[\begin{matrix}
   h(\rr)-\lambda &  \Delta(\rr)    \\
   \Delta^*(\rr)  &  h(\rr)-\lambda
   \end{matrix}\right]
   \left[\begin{matrix}
   U_k(\rr)    \\
   V_k(\rr)
   \end{matrix}\right]
   =
   E_k
   \left[\begin{matrix}
   U_k(r)    \\
   V_k(r)
   \end{matrix}\right],
                                                                  \label{eqn.01}
\end{equation}
where $h$ denotes the single-particle Hartree-Fock Hamiltonian;
$\lambda$ is the Fermi energy;
$\Delta$ is the pairing potential;
$U_k$ and $V_k$ are the upper and lower components of quasi-particle
wave functions respectively;
and $E_k$ denotes the quasi-particle energy.
 The HFB equation has been solved iteratively
and the modified Broyden method has been adopted for better convergence~\cite{hfbax}.

For the particle-hole interaction channel, the widely used Skyrme forces such as SkM$^*$~\cite{Bartel1982},
SLy4~\cite{Chabanat1998}, UNEDF0~\cite{unedf0} have been adopted. SkM$^*$ force has good surface properties
and has been widely applied in descriptions of fission. SLy4 force has been widely used in descriptions of neutron-rich nuclei.
UNEDF0 has been well optimized for descriptions of global binding energies. We speculate that a single density dependent term in standard Skyrme forces
is not sufficient for the Skyrme force to simulate  many-body correlations.
In addition, the extended
SkM$^*$$_{\rm ext1}$ and UNEDF0$_{\rm ext1}$~\cite{ZZW2018} forces
are also adopted.
Note that SkM$^*$$_{\rm ext1}$ and UNEDF0$_{\rm ext1}$ have an
additional high-order density-dependent term to improve global descriptions of binding energies.
In the particle-particle channel, a density-dependent pairing interaction
has been adopted as~\cite{wang},
\begin{equation}
    V_{pair}(\rr)
    =
    V_0
    \left\{1-\eta {\left[\frac{\rho(r)}{\rho_0(r)}\right]}^\gamma \right\},
                                                                  \label{eqn03}
\end{equation}
where $\rho_0(r)$ is the nuclear saturation density and
 $V_0,~\eta,~\gamma$ are three adjustable parameters.
With a pairing window of 60 MeV and the constants $\eta=0.8$, $\gamma=0.7$, we
adjusted pairing strength $V_0$ to reproduce the pairing gaps of $^{120}$Sn of $\Delta_n=1.245$ MeV.
The adopted pairing strengthes $V_0$ corresponding to SkM$^*$, SkM$^*$$_{\rm ext1}$, SLy4, UNEDF0, and UNEDF0$_{\rm ext1}$
are 386.0, 389.8, 449,7, 342.0, 339.6 MeV, respectively.
The pairing gaps could be very different towards drip lines by using different pairing interaction forms.
The resulted pairing gaps with Eq.(\ref{eqn03}) at the neutron drip lines are between the surface pairing and the mixed pairing\cite{wang}.
This is a reasonable choice because the pairing gaps  obtained  with the surface pairing interaction are too large toward the neutron drip lines if the pairing strength is invariant for stable and weakly bound nuclei.

%%%%%%%%%%%%%%%%%%%%%%%%%%%%%%%%%%%%%%%%%%%%%%%%%%%%%%%%%%%%%%%%%%%%%%%%%%%%%%%%
%123456789 123456789 123456789 123456789 123456789 123456789 123456789 123456789

To study odd-$A$ nuclei, the self-consistent quasiparticle blocking has been adopted.
For even-even nuclei, the particle density
$\rho(r)$ and the pairing density $\tilde{\rho}(r)$ are given as,
\begin{eqnarray}
   \rho(r)
   &=&
   2\sum_k  V_k^*(r) V_k(r),\nonumber\\
   \tilde{\rho}(r)
   &=&
   -2\sum_k  V_k(r) U_k^*(r),
                                                                  \label{eqn.04}
\end{eqnarray}
where the quasiparticle energy cutoff is taken as 60 MeV.
For odd-$A$ nuclei, by identifying the blocked quasiparticle state $\mu$, the particle density
and the pairing density are given as~\cite{xiong,Bertsch2009},
\begin{eqnarray}
   \rho_{\rm odd}(\rr)
   &=&
   |U_\mu(\rr)|^2 + |V_\mu(\rr)|^2
   + 2\sum_{k\neq \mu}  |V_k(r)|^2 ,\nonumber\\
   \tilde{\rho}_{\rm odd} (r)
   &=&
   -2\sum_{k\neq \mu}  V_k(r) U_k^*(r).
                                                                  \label{eqn.05}
\end{eqnarray}
In this case, the blocked state doesn't contribute to the pairing density.
The particle number equation has to be modified accordingly.
The blocking calculations are performed only for odd-$Z$ and even-$N$ nuclei, in which
the blocked proton states are discrete and they can be identified clearly.

To study the deformation properties, we also performed deformation constrained calculations of
potential energy curves. The shape coexistence is possible around $^{40}$Mg.
For better numerical convergence of the constrained calculations, the  augmented Lagrangian method
has been adopted~\cite{ALM}.

\section{Results and discussions}
%%%%%%%%%%%%%%%%%%%%%%%%%%%%%%%%%%%%%%%%%%%%%%%%%%%%%%%%%%%%%%%%%%%%%%%%%%%%%%%%
%%%%%%%%%%%%%%%%%%%%%%%%%%%%%%%%%%%%%%%%%%%%%%%%%%%%%%%%%%%%%%%%%%%%%%%%%%%%%%%%
\begin{figure}[htbp]
\centering
\includegraphics[width=0.55\textwidth]{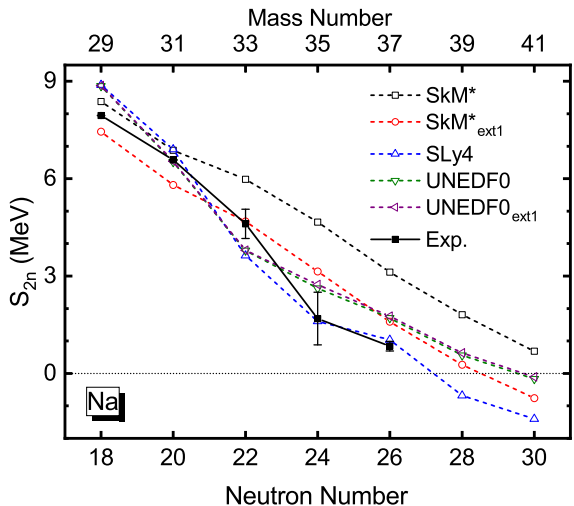}
\caption{
(Color online)
The two-neutron separation energies  $S_{2n}$  of sodium isotopes obtained by HFB calculations with SkM$^*$~\cite{Bartel1982},
SLy4~\cite{Chabanat1998}, UNEDF0~\cite{unedf0},
SkM$^*$$_{\rm ext1}$~\cite{ZZW2018}, and UNEDF0$_{\rm ext1}$~\cite{ZZW2018} forces.
Experimental data taken from~\cite{ame2016} are also shown.
                                                                   \label{FIG1}
}
\end{figure}
%description for Fig.1

Firstly, we studied the two-neutron separation energies  $S_{2n}$ of sodium isotopes to the drip-line with
various Skyrme forces.
The recent experiment performed in RIKEN has newly observed one event of $^{39}$Na.
This has not been observed before, indicting that $^{39}$Na is very weakly bound and is most likely to be the drip-line.
Fig.\ref{FIG1} displays the calculated two-neutron separation energies
with SkM$^*$, SLy4, UNEDF0, SkM$^*$$_{\rm ext1}$ and UNEDF0$_{\rm ext1}$
forces.
Note that the experimental $S_{2n}$ of $^{35}$Na and $^{37}$Na are actually taken from the estimations
of AME2016~\cite{ame2016}.
In calculations with SkM$^*$, $^{41}$Na is the drip line. Based on systemic calculations~\cite{ZZW2018},
SkM$^*$ overestimates binding energies towards the neutron drip-line.
In SLy4 calculations,  $^{37}$Na is the drip-line nucleus. The UNEDF0 and UNEDF0$_{\rm ext1}$ calculations
obtained very similar results. We see that in calculations with UNEDF0, UNEDF0$_{\rm ext1}$ and SkM$^*$$_{\rm ext1}$
forces, $^{39}$Na is the drip-line of sodium isotopes.
Therefore these calculations provide a  stringent  constraint on the drip-line. The reasonable drip-line should be located between SLy4 and SkM$^*$ results.
Consequently, we adopt SkM$^*$$_{\rm ext1}$ and  UNEDF0$_{\rm ext1}$ forces in systematic calculations of drip-line nuclei.

%%%%%%%%%%%%%%%%%%%%%%%%%%%%%%%%%%%%%%%%%%%%%%%%%%%%%%%%%%%%%%%%%%%%%%%%%%%%%%%%

%%%%%%%%%%%%%%%%%%%%%%%%%%%%%%%%%%%%%%%%%%%%%%%%%%%%%%%%%%%%%%%%%%%%%%%%%%%%%%%%
\begin{figure}[htbp]
\centering
\includegraphics[width=0.5\textwidth]{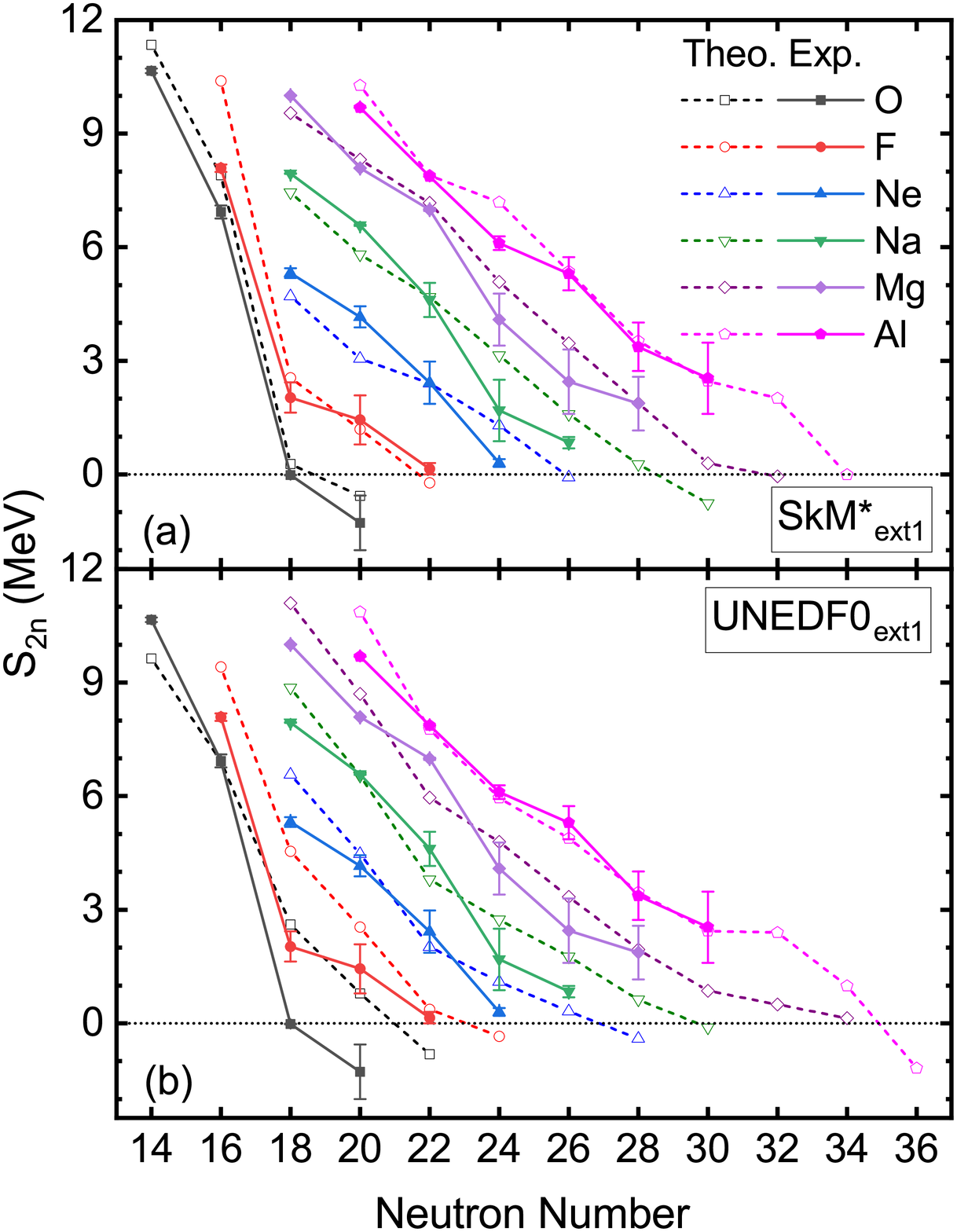}
\caption{
(Color online)
The calculated two-neutron separation energies  $S_{2n}$ of oxygen, fluorine, neon,
sodium, magnesium and aluminum isotopes with the SkM$^*$$_{\rm ext1}$ and UNEDF0$_{\rm ext1}$ forces.
The corresponding experimental data from~\cite{ame2016}  are also shown.
                                                                   \label{FIG2}
}
\end{figure}
%%%%%%%%%%%%%%%%%%%%%%%%%%%%%%%%%%%%%%%%%%%%%%%%%%%%%%%%%%%%%%%%%%%%%%%%%%%%%%%%
%description for Fig.2

Figure~\ref{FIG2} shows the systematic two-neutron separation energies of O, F, Ne, Na, Mg, Al isotopes.
The calculations are performed for even-$N$ isotopes since the pairing can enhance the stabilities of weakly bound nuclei.
The experimental drip lines of oxygen and fluorine are $^{24}$O and $^{31}$F respectively.
There is a surprising gap of 6 neutrons from O to F drip-line locations, presenting a theoretical challenge.
In our calculations, we see that $S_{2n}$ of O and F isotopes can be nicely reproduced by SkM$^*$$_{\rm ext1}$.
$^{26}$O is slightly bounded by 270 keV and  $^{31}$F is slightly unbound by 220 keV.
This is close to the experimental data with deviations about 300 keV.
In the literature, the drip-line of oxygen has been related
to the roles of tensor forces and three-body forces~\cite{otsuka}.
It is still challenging to reproduce simultaneously the drip lines of oxygen and fluorine isotopes.
Our results can almost resolve the puzzle of O-F drip lines.
We demonstrated that the successful HFB calculations have largely reduced the roles of additional effective forces at the ground-state level.
However, the UNEDF0$_{\rm ext1}$ results are not satisfactory for oxygen and fluorine isotopes although
they are much better in global calculations  of binding energies especially in the heavy-mass region than SkM$^*$$_{\rm ext1}$ results.

The overall $S_{2n}$ of neon and sodium isotopes can not be nicely reproduced compared to oxygen and fluorine isotopes.
There are deformation transitions due to the onset of ``island  of inversion" around $^{32}$Mg so that Skyrme forces have not been optimized so well.
Nevertheless, the experimental neon drip-line is at $^{34}$Ne which can be reproduced by SkM$^*$$_{\rm ext1}$.
The UNEDF0$_{\rm ext1}$ has slightly overestimated the neon drip-line.

The existence of $^{39}$Na has provided a critical examination of various Skyrme forces.
In Ref.~\cite{Erler2012nature} by Erler {\it et al.},  generally SkM$^*$ produces the furthest neutron drip lines while SLy4 produces
the nearest neutron drip lines. These two extreme cases both have been excluded by the drip-line nuclei of $^{34}$Ne and $^{39}$Na.
The $S_{2n}$ of $^{39}$Na obtained by UNEDF0$_{\rm ext1}$ and SkM$^*$$_{\rm ext1}$ are about 0.63 and 0.27 MeV, respectively.
Associated with this subtle difference, the neutron drip lines predicted by UNEDF0$_{\rm ext1}$ are systematically farther than that by SkM$^*$$_{\rm ext1}$
for Ne, Mg, Al isotopes. Thus the future measurement of $S_{2n}$ of $^{39}$Na will also be a critical constraint.

For Mg isotopes, the ground state and first excited state of $^{40}$Mg has been observed~\cite{Baumann2007,crawford}.
In our calculations,  $S_{2n}$ of $^{40}$Mg are 1.88 MeV by SkM$^*$$_{\rm ext1}$ and 1.95 MeV by  UNEDF0$_{\rm ext1}$, which are consistent with the AME2016 estimations~\cite{ame2016}.
The first excited state is around 500 keV~\cite{crawford}, which is below the $S_{2n}$ threshold.
With  SkM$^*$$_{\rm ext1}$, the $S_{2n}$ of $^{42}$Mg and $^{44}$Mg are 0.29 MeV and $-$0.05 MeV, respectively.
With UNEDF0$_{\rm ext1}$, the $S_{2n}$ of $^{42}$Mg and $^{44}$Mg   are 0.86 MeV and 0.49 MeV, respectively.
Indeed, UNEDF0$_{\rm ext1}$  predicts slightly further neutron drip lines compared to SkM$^*$$_{\rm ext1}$
as also shown in sodium and neon isotopes.
Based on the fact that $^{39}$Na is weakly bound, thus we can confidently infer that $^{42}$Mg is a weakly bound nucleus
based on both SkM$^*$$_{\rm ext1}$ and  UNEDF0$_{\rm ext1}$ results.
There is a possibility that  $^{44}$Mg is very weakly bound but currently it can not be verified.

For Al isotopes, the last observed isotope is $^{43}$Al~\cite{Baumann2007}. Both calculations with SkM$^*$$_{\rm ext1}$ and UNEDF0$_{\rm ext1}$
 can nicely reproduce  $S_{2n}$ of Al isotopes. The obtained $S_{2n}$ of $^{45}$Al is 2.0 MeV by SkM$^*$$_{\rm ext1}$
and 2.4 MeV by  UNEDF0$_{\rm ext1}$. Therefore, we are confident that $^{45}$Al is bound and more stable than $^{42}$Mg.
As for $^{47}$Al, its $S_{2n}$ decrease very rapidly in both calculations. The obtained $S_{2n}$ of $^{47}$Al is $-$0.01 MeV by SkM$^*$$_{\rm ext1}$
and 0.99 MeV by  UNEDF0$_{\rm ext1}$.  Therefore, it is difficult to conclude that $^{47}$Al is weakly bound or not bound.

For other mass models, the drip lines from the Weizs\"{a}cker-Skyrme formula~\cite{ws4} are at $^{29}$F, $^{34}$Ne,  $^{37}$Na,  $^{40}$Mg and $^{43}$Al respectively, which
slightly underestimated the neutron drip-line.
In Gogny-HFB calculations with the D1S force~\cite{gogny}, the drip lines are at  $^{29}$F, $^{32}$Ne,  $^{35}$Na,  $^{40}$Mg and $^{43}$Al respectively, which
has systematically underestimated the neutron drip-line.
In the HFB-21 mass table~\cite{hfb21}, the drip-line nuclei are   $^{29}$F ($S_{2n}$ =1.37 MeV), $^{34}$Ne (0.16 MeV),  $^{37}$Na (1.0 MeV),  $^{42}$Mg (0.56 MeV) and $^{45}$Al (0.91 MeV) respectively, which
 slightly underestimated the neutron drip-line and is satisfactory compared to SkM$^*$$_{\rm ext1}$ results.
In the Relativistic-Hartree-Bogoliubov calculations with the PC-PK1 functional~\cite{xia},  the drip lines are at   $^{31}$F, $^{42}$Ne,  $^{45}$Na,  $^{46}$Mg and $^{49}$Al respectively, which
has much overestimated the neutron drip-line.
In the macro-microscopic model calculations~\cite{moller}, the drip lines are at  $^{31}$F, $^{34}$Ne,  $^{39}$Na,  $^{40}$Mg and $^{41}$Al respectively, which
has underestimated the neutron drip-line around $^{41}$Al.
By our calculations and examinations of $^{39}$Na, the constraint on neutron drip-line has been much improved.
We see that the overall performance of SkM$^*$$_{\rm ext1}$ is the best to predict the drip lines from oxygen to aluminum isotopes.
Based on both SkM$^*$$_{\rm ext1}$ and UNEDF0$_{\rm ext1}$ calculations,  we can infer that $^{42}$Mg is weakly bound and $^{45}$Al is less weakly bound, and $^{44}$Mg and $^{47}$Al could be barely existed.
In the latest {\it  ab initio} calculations~\cite{holt}, the probabilities to bound of $^{42}$Mg and
$^{44}$Mg are 78$\%$ and 41$\%$ respectively.  The probabilities to bound of $^{45}$Al and
$^{47}$Mg are 91$\%$ and 59$\%$ respectively. The {\it  ab initio}  results are well consistent with both SkM$^*$$_{\rm ext1}$ and UNEDF0$_{\rm ext1}$ results within uncertainties.

%%%%%%%%%%%%%%%%%%%%%%%%%%%%%%%%%%%%%%%%%%%%%%%%%%%%%%%%%%%%%%%%%%%%%%%%%%%%%%%%

%%%%%%%%%%%%%%%%%%%%%%%%%%%%%%%%%%%%%%%%%%%%%%%%%%%%%%%%%%%%%%%%%%%%%%%%%%%%%%%%
\begin{figure}[htbp]
\centering
\includegraphics[width=0.5\textwidth]{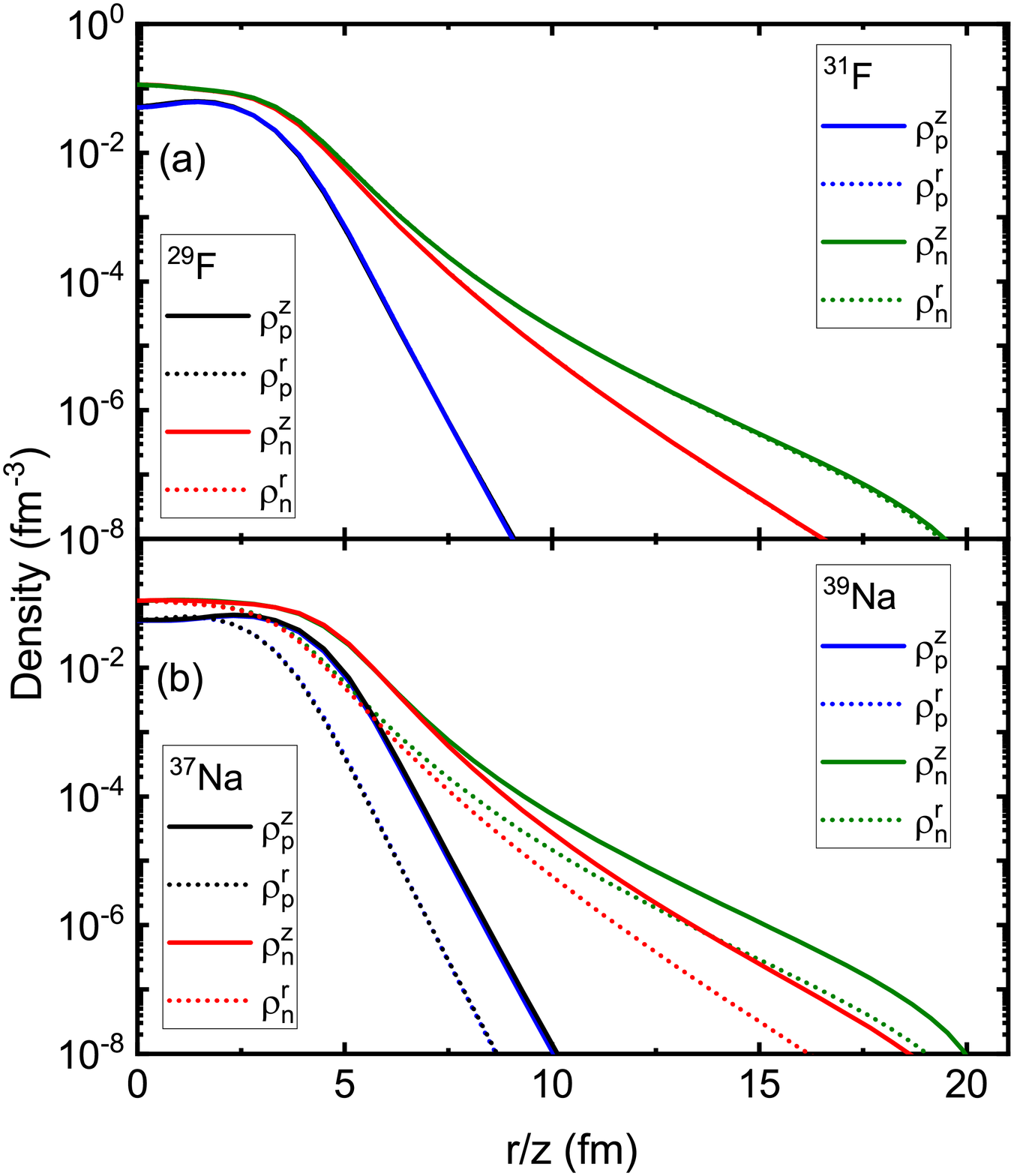}
\caption{
(Color online)
The proton and neutron density distributions of $^{29,31}$F and $^{37,39}$Na calculated with the SkM$^*$$_{\rm ext1}$ force.
The density distributions are presented in the cylindrical coordinate
spaces as $\rho(r,z)$. The density profiles along $r$-axis (dotted lines) and $z$-axis  (solid lines) are
shown respectively.
                                                                   \label{FIG3}
}
\end{figure}
%%%%%%%%%%%%%%%%%%%%%%%%%%%%%%%%%%%%%%%%%%%%%%%%%%%%%%%%%%%%%%%%%%%%%%%%%%%%%%%%
%description for Fig.3

Figure \ref{FIG3} shows the density distributions of $^{29,31}$F and $^{37,39}$Na.
The densities are obtained by HFB calculations with the  SkM$^*$$_{\rm ext1}$ force.
$^{31}$F is slightly unbound in terms of $S_{2n}$ while its neutron Fermi energy $\lambda_n$
is $-$0.16 MeV. $^{31}$F is spherical in both calculations.
Compared to $^{29}$F, we see that $^{31}$F ($N$=22) has a spherical neutron halo structure at outer surfaces.
This is consistent with the recent Gamow shell model studies of $^{31}$F~\cite{lijg}.
In our calculations,  $^{39}$Na is prolate deformed with a quardpole deformation $\beta_2$=0.35.
With blocking and noblocking calculations, the binding energies are 242.72 MeV and 243.03 MeV, respectively, which are close.
The profiles of density distributions in cylindrical coordinate spaces are displayed along the transverse $r$-axis and the principle $z$-axis respectively.
It can be seen that $^{39}$Na has a deformed halo structure.
In both $^{31}$F and  $^{39}$Na, the halo structures are mainly related to continuum contributions.
In calculations with UNEDF0$_{\rm ext1}$, the halo structure is less prominent due to its larger  $S_{2n}$ compared
to SkM$^*$$_{\rm ext1}$ results.

%%%%%%%%%%%%%%%%%%%%%%%%%%%%%%%%%%%%%%%%%%%%%%%%%%%%%%%%%%%%%%%%%%%%%%%%%%%%%%%%

%%%%%%%%%%%%%%%%%%%%%%%%%%%%%%%%%%%%%%%%%%%%%%%%%%%%%%%%%%%%%%%%%%%%%%%%%%%%%%%%
%%%%%%%%%%%%%%%%%%%%%%%%%%%%%%%%%%%%%%%%%%%%%%%%%%%%%%%%%%%%%%%%%%%%%%%%%%%%%%%%
%%%%%%%%%%%%%%%%%%%%%%%%%%%%%%%%%%%%%%%%%%%%%%%%%%%%%%%%%%%%%%%%%%%%%%%%%%%%%%%%
\begin{figure}[htbp]
\centering
\includegraphics[width=0.5\textwidth]{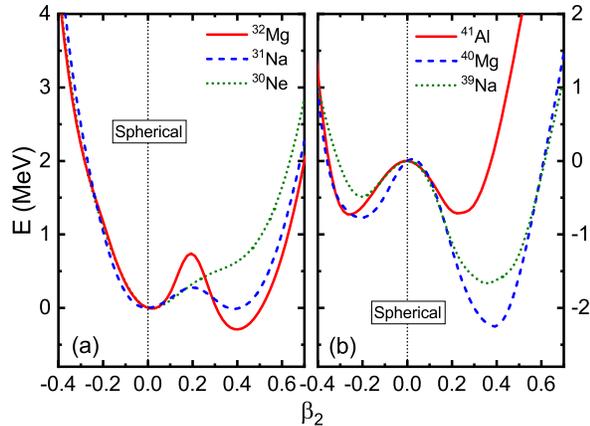}
\caption{
(Color online)
The calculated potential energy curves of $N$=20 and $N$=28 drip-line nuclei a function of quardruple
deformations, with the SkM$^*$$_{\rm ext1}$ force.
                                                                   \label{FIG4}
}
\end{figure}

Figure \ref{FIG4} shows the potential energy curves of $N$=20 and $N$=28 drip-line nuclei as a function of quardruple
deformations. Due to the breaking of magic neutron numbers in the neutron-rich region, there are possibly
rich deformations coexist. The calculations are performed without quasiparticle blocking since it is rather complicated due to shape transitions.
It is interest to study the deformation properties of the weakly bound $^{39}$Na.
The deformations of $^{32}$Mg are also of critical interests since it is at the center of
the ``island of inversion". In calculations with SkM$^*$$_{\rm ext1}$, we see that $^{32}$Mg
has a prolate ground state ($\beta_2$=0.39) and a second energy minimum at the spherical shape.
The energy of the prolate shape is lower than the spherical case by 0.29 MeV.
While experimentally the second $0^{+}$ in $^{32}$Mg has an energy of 1.058 MeV~\cite{mg32}.
The ground state has a large prolate deformation indicate that SkM$^*$$_{\rm ext1}$
can actually reproduce the deformation of ``island of inversion", which has been a longstanding issue of mean-field calculations,
although its $0_2^{+}$ energy is not perfectly reproduced and beyond mean-field effects should be taken into account.
In our calculations with SkM$^*$$_{\rm ext1}$, $^{31}$Na has a prolate-spherical shape coexistence, in which energies of coexisting shapes are almost the same.
Following this shape evolution,  $^{30}$Ne has a  spherical ground state and $^{29}$F is unambiguously spherical.
Note that  $^{30}$Ne is prolately deformed in experiments~\cite{Michimasa2014}.
This is related to that  SkM$^*$$_{\rm ext1}$ can reproduce the prolate deformation of $^{32}$Mg but the inversion energy is not so large yet.
On the other hand,  calculations with UNEDF0$_{\rm ext1}$ resulted in a spherical
ground state for $^{32}$Mg.
For $^{39}$Na, we see that its ground state has a large prolate deformation and the second
energy minimum has an oblate shape. The prolate-oblate shape coexistence in $^{39}$Na is very similar to that of $^{40}$Mg~\cite{Terasaki1997}.
The detailed studies of shape transitions are important to precisely calculate two-neutron
separation energies in this region. For $^{41}$Al, the oblate shape has an energy of 18 keV lower than that of the prolate shape.
Therefore, it is hard to identify the ground state deformation of $^{41}$Al and it is the novel crossover of the shape transition with two $0^{+}$ states with close energies, while $^{42}$Si has an unambiguous oblate ground state~\cite{42si}.
This is consistent with  Gogny-HFB calculations that $^{41}$Al has close energies in oblate and prolate shapes~\cite{gogny}.

%%%%%%%%%%%%%%%%%%%%%%%%%%%%%%%%%%%%%%%%%%%%%%%%%%%%%%%%%%%%%%%%%%%%%%%%%%%%%%%%

%%%%%%%%%%%%%%%%%%%%%%%%%%%%%%%%%%%%%%%%%%%%%%%%%%%%%%%%%%%%%%%%%%%%%%%%%%%%%%%%
\section{summary}
%%%%%%%%%%%%%%%%%%%%%%%%%%%%%%%%%%%%%%%%%%%%%%%%%%%%%%%%%%%%%%%%%%%%%%%%%%%%%%%%

In summary, we studied the ground-state properties of light nuclei from oxygen to aluminum isotopes around the newly
observed $^{39}$Na with the Skyrme-Hartree-Fock-Bogoliubov framework in the deformed coordinate spaces.
Based on the fact that  $^{39}$Na is a weakly bound drip-line nucleus, we
have carefully examined various theoretical models. This provides
a critical theoretical constraint on predictions of neutron drip lines.
As a result, the SkM$^*$$_{\rm ext1}$ parameterization is demonstrated to  be
a very good choice for  drip-line nuclei.
It can nicely reproduce the trend in two-neutron separation energies of oxygen and fluorine isotopes.
Furthermore, it resulted in a large prolate deformation in the ground state of $^{32}$Mg,
which is at the center of the ``island of inversion", although the inversion energy is not so large.
Based on comprehensive examinations, we can infer that $^{42}$Mg is weakly bound and $^{45}$Al is less weakly bound,
while $^{44}$Mg and $^{47}$Al could be barely existed.
Our calculations shows a deformed halo structure in $^{39}$Na. The shape transitions in neighboring nuclei
indicates that $^{41}$Al is the novel crossover with almost the same energies in oblate and prolate shapes.
Definitely more experimental information in this region can further
improve predictions for ground states and excited states. We hope our studies will be valuable for experiments
performed in such as the forthcoming FRIB and other RIB facilities.

%%%%%%%%%%%%%%%%%%%%%%%%%%%%%%%%%%%%%%%%%%%%%%%%%%%%%%%%%%%%%%%%%%%%%%%%%%%%%%%%
\acknowledgments
 This work was supported by the National Natural Science Foundation of China under Grants No. 11975032, 11835001, 11790325;  the
 National Key R$\&$D Program of China (Contract No. 2018YFA0404403).
We also acknowledge that computations in this work were performed in Tianhe-1A
located in Tianjin.

\end{document}